\documentclass[aps,twocolumn,showpacs,superscriptaddress,citeautoscript,reprint]{revtex4-1}

\usepackage{bm}
\usepackage{graphicx}
\usepackage{amsmath}
\usepackage{amsfonts}
\usepackage{MnSymbol}

\newcommand{\rmi}{\mathrm{i}\,}

\newcommand{\Tr}{\mathrm{Tr}\,}

\renewcommand{\Re}{\mathrm{Re}\,}

\renewcommand{\Im}{\mathrm{Im}\,}

\begin{document}

\title{Interactions and thermoelectric effects in a parallel-coupled double quantum dot}

\author{Miguel A.\ Sierra}

\affiliation{Instituto de F\'{\i}sica Interdisciplinar y Sistemas Complejos IFISC (UIB-CSIC), E-07122 Palma de Mallorca, Spain}

\author{M.\ Saiz-Bret\'{i}n}

\affiliation{GISC, Departamento de F\'{\i}sica de Materiales, Universidad Complutense, E-28040 Madrid, Spain}

\affiliation{Department of Physics, University of Warwick, Coventry, CV4 7AL, United Kingdom}

\author{F. Dom\'{\i}nguez-Adame}

\affiliation{GISC, Departamento de F\'{\i}sica de Materiales, Universidad Complutense, E-28040 Madrid, Spain}

\affiliation{Department of Physics, University of Warwick, Coventry, CV4 7AL, United Kingdom}

\author{David\ S\'{a}nchez}

\affiliation{Instituto de F\'{\i}sica Interdisciplinar y Sistemas Complejos IFISC (UIB-CSIC), E-07122 Palma de Mallorca, Spain}


\begin{abstract}

We investigate the nonequilibrium transport properties of a double quantum 
dot system connected in parallel to two leads, including intradot 
electron-electron interaction. In the absence of interactions the system 
supports a bound state in the continuum. This state is revealed as a Fano 
antiresonance in the transmission when the energy levels of the dots are 
detuned. Using the Keldysh nonequilibrium Green's function formalism, we find 
that the occurrence of the Fano antiresonance survives in the presence of 
Coulomb repulsion. We give precise predictions for the experimental detection 
of bound states in the continuum. First, we calculate the differential 
conductance as a function of the applied voltage and the dot level detuning 
and find that crossing points in the diamond structure are revealed as minima 
due to the transmission antiresonances. Second, we determine the 
thermoelectric current in response to an applied temperature bias. In the 
linear regime, quantum interference gives rise to sharp peaks in the 
thermoelectric conductance. Remarkably, we find interaction induced strong 
current nonlinearities for large thermal gradients that may lead to several 
nontrivial zeros in the thermocurrent. The latter property is especially 
attractive for thermoelectric applications.
  
\end{abstract}

\maketitle

\section{Introduction}

Double quantum dots (DQDs) coupled in parallel serve as an excellent platform 
to test interaction and interference effects~\cite{hol01,hol02,che04} 
because, due to the small system size, Coulomb repulsion is intensified and 
quantum phase-coherence is preserved during an electron transfer between the 
attached reservoirs. Destructive interference between different paths 
traversing the two dots can lead to Fano resonances~\cite{lad03,lu05,tro07} 
and Dicke effect in the presence of a magnetic flux~\cite{ore04}. The 
interference pattern can be ultimately influenced by Coulomb 
interactions~\cite{kon02,din05,tan05,lu06c}. In the case of strong 
correlations, there may arise competition between magnetic interactions and 
Kondo effect~\cite{lop01,krycho075}, Kondo states with higher 
symmetry~\cite{Litvin245,lop05} and signatures of quantum phase 
transitions~\cite{won12}. Furthermore, these systems are interesting for 
solid-state quantum bit implementations and as probes of 
entanglement~\cite{los00,ram06}.

Consider for the moment the case without interactions. Fano effect originates 
from the quantum interference between resonant and nonresonant 
processes~\cite{Fano61} giving rise to asymmetric electric conductance 
profiles in the nanostructure~\cite{mir10}. The occurrence of Fano 
antiresonances in DQD systems traces back to bound states in the continuum 
(BIC), formerly introduced in quantum mechanics by von Neumann and Wigner in 
1929 as squared-integrable solutions buried in the continuum energy 
spectrum~\cite{Neumann29}. In the absence of interdot tunnel coupling, the 
local density of states at the quantum dots presents a narrow peak that 
approaches a $\delta$-function when the coupling to the leads vanishes, 
indicating that the resonant level becomes a truly 
BIC~\cite{Gonzalez-Santander13,Alvarez15}. It should be stressed that the 
interdot tunneling should be kept as small as possible to detect to 
occurrence of BICs in nonequilibrium transport experiments~\cite{Alvarez15}. 
In this work, we would first like to analyze to what extent BIC survive in 
the presence of Coulomb interaction. This is a natural question since small 
dots exhibit in experiments Coulomb blockade effects due to a finite on-site 
charging energy, which is typically the largest energy scale of the problem. 
Below we demonstrate that BICs are robust against charging effects and that, 
in fact, a replica of the original BIC emerges at high energies due to Coulomb 
repulsion.

The second objective of this paper is to address the question as to whether 
parallel coupled DQDs are useful in thermoelectrics. The problem has thus far 
attracted little attention~\cite{Liu10,Trocha12} despite the fact that sharp 
resonances (such as the Fano lineshapes) are in principle ideal candidates 
for highly efficient waste heat-to-electricity converters~\cite{mahan}. The 
issue can be viewed within the broader perspective of nanostructures as key 
elements in future thermoelectric applications based on the Seebeck effect. 
Such devices have many attractive features compared with other methods due to 
the absence of moving parts, scalability and high 
reliability~\cite{Goldsmid10}. Closely connected with many discoveries that 
have demonstrated that nanometer-sized objects exhibit physical properties 
not shared by bulk materials~\cite{san14}, theoretical 
predictions~\cite{Hicks93,Khitun00,Balandin03} and 
experiments~\cite{Venkata01,Harman02,Hochbaum08,Boukai08} have pointed out 
that thermoelectric properties at the nanoscale can be strongly enhanced. In 
fact, an enhancement of the thermoelectric figure of merit can be achieved 
with the aid of quantum interference 
phenomena~\cite{Bergfield09,Karlstrom11,Saiz-Bretin15}. In particular, 
quantum effects giving rise to Fano antiresonances in the transmission 
probability of DQD artificial molecules lead to departures from the 
Wiedemann-Franz law~\cite{Gomez-Silva12,Wojc085}. As a consequence, the 
thermoelectric efficiency is greatly enhanced~\cite{Zheng12}. While 
Refs.~\cite{Liu10,Trocha12} neglect the role of BICs and focus on the linear 
regime of transport, here we find that the generated thermocurrent in 
response to an applied temperature bias can be large for small thermal 
gradients due to the important role of BICs in the DQD spectral density. More 
importantly, we obtain strongly nonlinear current-temperature characteristics 
that present different nontrivial zeros with enhanced peak-to-valley ratios. 
This is an appealing functionality that relies on the electron-electron 
interaction in the DQD system and disappears in the noninteracting limit. 

Our work is structured as follows. In Sec.~II we discuss our theoretical 
model for a parallel coupled DQD system with negligible 
interdot tunneling, but finite intradot Coulomb interaction. We calculate the 
electric current using the Keldysh nonequilibrium Green's function 
formalism~\cite{Haug08}. Specifically, we consider the Coulomb blockade 
regime and disregard cotunneling and Kondo correlations within an 
equation-of-motion (EOM) technique. In Sec.~III we examine the spectral function 
and explicitly show the emergence of a BIC as a function of the detuning of 
the dot levels and the couplings between the dots and the leads. We here 
compare the cases with and without electron-electron interactions. The 
transmission probability is discussed in Sec.~IV. Since current can be driven 
by either electric or thermal means we first treat the voltage transport in 
Sec.~V and then the thermoelectric transport in Sec.~VI. In both cases we 
consider the linear conductances and the nonlinear regime of transport. 
Finally, Sec.~VII summarizes our conclusions.

\section{Model and formalism}

We consider two quantum dots forming a DQD system, connected to left~(L) and 
rightł(R) leads by tunnel couplings, as shown schematically in 
Fig.~\ref{fig1}. Only one energy level in each dot is assumed relevant and 
electron-phonon interaction is neglected hereafter. However, onsite Coulomb interaction is considered for each dot.

\subsection{System Hamiltonian}

\begin{figure}[t]
\begin{center}
\includegraphics[width=0.8\linewidth]{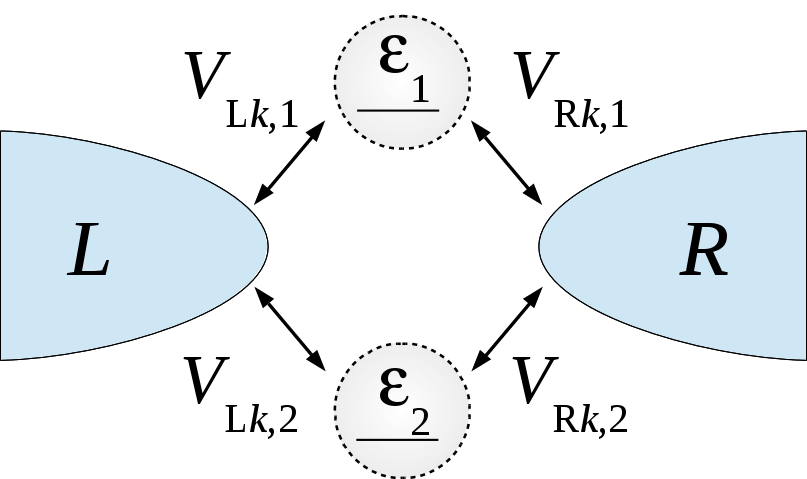}
\end{center}
\caption{(Color online) Schematic diagram of the DQD system, connected to leads $L$ and $R$. We indicate the energy levels $\varepsilon_1$ and $\varepsilon_2$, and the tunnel amplitudes $V_{Lk,1},\;V_{Lk,2},\;V_{Rk,1}$  and $V_{Rk,2}$. We take into account intradot electron-electron interactions (not shown here).}
\label{fig1}
\end{figure}

The Hamiltonian associated to the whole system can be written as
$\mathcal{H}=\mathcal{H}_\mathrm{DQD}+\mathcal{H}_\mathrm{leads}+\mathcal{H}_\mathrm{tunnel}$.
Here, $\mathcal{H}_\mathrm{DQD}$ describes the dynamics of interacting electrons in the DQD system
\begin{subequations}
\begin{equation}
\mathcal{H}_\mathrm{DQD}=\sum_{i\sigma} \varepsilon_i^{}n_{i\sigma} +\sum_i U_i n_{i\uparrow}n_{i\downarrow}\ ,
\label{eq:01a}
\end{equation}
where the index $i=1,2$ runs over the quantum dots and $U_i$ is the Coulomb energy when the dots are doubly occupied. The number operator is $n_{i\sigma}=d_{i\sigma}^{\dag}d_{i\sigma}^{}$, where $d_{i\sigma}^{\dag}$ ($d_{i\sigma}^{}$) is the creation (annihilation) operator of an electron in the dot $i$ with energy $\varepsilon_{i}$ and spin $\sigma$. Quantum dots are assumed to be far apart so tunnel and capacitive couplings between them are weak. This weakness facilitates the possibility of detecting the BICs in transport experiments~\cite{Alvarez15}. Electrons in the ideal leads are regarded as noninteracting particles with crystal momentum $k$ and energy $\varepsilon_{\alpha k \sigma}$. The corresponding Hamiltonian is
\begin{equation}
\mathcal{H}_\mathrm{leads}=\sum_{\alpha k \sigma} \varepsilon_{\alpha k \sigma}^{} C_{\alpha k \sigma}^\dag C_{\alpha k \sigma}^{}\ .
\label{eq:01b}
\end{equation}
Here, $C_{\alpha k\sigma}^{\dag}$ ($C_{\alpha k\sigma}^{}$) denotes the creation (annihilation) operator of a conduction electron in the semi-infinite lead $\alpha=L,R$.  Finally, the quantum dots are tunnel coupled to both leads (see Fig.~\ref{fig1}). Hence,
\begin{equation}
\mathcal{H}_\mathrm{tunnel}=\sum_{\alpha k \sigma i} V_{\alpha k,i}^{} C_{\alpha k \sigma}^{\dag} d_{i\sigma}^{} + \textrm{H.c.}\ , 
\label{eq:01c}
\end{equation}
\label{eq:01}
\end{subequations}
where $\textrm{H.c.}$ stands for Hermitian conjugate and the amplitudes $V_{\alpha k,i}$ are tunnel coupling parameters. These factors are spin-independent for nonmagnetic leads. 

\subsection{Charge current}

The charge current is obtained as the time derivative of the expected occupation in one of the leads, $I_{\alpha}=-ed\langle n_\alpha(t)\rangle /dt$ with $n_\alpha(t)=\sum_{k\sigma}C_{\alpha k \sigma}^\dag(t) C_{\alpha k \sigma}^{}(t)$. In the steady state case, charge conservation demands that $I_L+I_R=0$. Hence we can define the charge current flowing through the DQD system as $I\equiv I_L=-I_R$. Following Ref.~\onlinecite{Haug08}, the charge current can be cast in the form (we set $\hbar=1$ hereafter)
\begin{subequations}
\begin{equation}
I=\frac{e}{\pi} \int_{-\infty}^\infty dE \sum_{k\sigma i}\Re{\left[V_{\alpha k,i}^{}G_{i\sigma,\alpha k\sigma}^{<}(E)\right]}\ ,
\label{eq:02a}
\end{equation}
where $G_{i\sigma,\alpha k\sigma}^{<}(E)$ is the Fourier transform of the lesser Green's function $G_{i\sigma,\alpha k\sigma}^{<}(t)\equiv \rmi \langle C_{\alpha k \sigma}^{\dag}(0) d_{i\sigma}(t)\rangle$.  

We now define the parameters $\Gamma^{\alpha}_{ij}=2\pi \rho_\alpha V_{\alpha k,i}^{}V_{\alpha k,j}^{*}$, $\rho_\alpha$ being the density of states of the lead $\alpha$. In the wide band limit these parameters are assumed to be independent of the electron energy. As a consequence, the numbers $\Gamma^\alpha_{ij}$ are constants. Langreth rules~\cite{Haug08} allow the charge current to be expressed solely in terms of the advanced and retarded Green's functions of the dots
\begin{eqnarray}
I&=&\frac{e}{2\pi}\sum_{\sigma}\int_{-\infty}^\infty dE\, \Big[f_L(E,T_L)-f_R(E,T_R)\Big]\nonumber \\
&\times & \Tr \Big[ {\bm G}^{a}_{\sigma\sigma}(E){\bm \Gamma}^{R}{\bm G}^{r}_{\sigma\sigma}(E){\bm \Gamma}^{L}\Big]\ ,
\label{eq:02b}
\end{eqnarray}
\label{eq:02}
\end{subequations}
where $f_{\alpha}(E,T_\alpha)=\left\{\exp\left[(E-\mu_\alpha)/k_BT_\alpha\right]+1\right\}^{-1}$ is the Fermi distribution function of the lead $\alpha$ with electrochemical potentials $\mu_L=\varepsilon_F+eV/2$ and $\mu_R=\varepsilon_F-eV/2$ and temperatures $T_L$ and $T_R$. The matrix elements of ${\bm \Gamma}^{\alpha}$ are given by $\Gamma^{\alpha}_{ij}$. In this expression ${\bm G}^{r}_{\sigma\sigma}(E)$ is a $2\times 2$ matrix whose elements are the Fourier transform of the
retarded Green's functions
\begin{equation}
G^{r}_{i\sigma,j\sigma}(t)=-\rmi\theta(t)\langle[d_{i\sigma}^{}(t),d_{j\sigma}^{\dag}(0)]_{+}\rangle\ ,
\label{eq:03}
\end{equation}
where $\theta$ is the Heaviside step function and $[\cdots]_{+}$ stands for the anticommutator. Similarly, $G^{a}_{i\sigma,j\sigma}(E)$ is the Fourier transform of the advanced Green's functions $G^{a}_{i\sigma,j\sigma}(t)=\rmi\theta(-t)\langle[d_{i\sigma}^{}(t), d_{j\sigma}^{\dag}(0)]_{+}\rangle$.

\subsection{Equation-of-motion approach} \label{mfa}

We restrict ourselves to the Coulomb-blockade regime, for which the Coulomb energy $U_i$
is assumed to be much larger than the background temperature $k_BT$ and the hybridation parameters $\Gamma^{\alpha}_{ij}$.  Using the EOM method, the retarded Green's functions can be assessed by neglecting the correlators $\llangle d_{i\sigma}^{} d_{i\bar{\sigma}}^{\dag} C_{\alpha k\bar{\sigma}}^{},d_{j\sigma}^{}\rrangle_t \simeq 0$ and $\llangle d_{i\sigma}^{}C_{\alpha k \bar{\sigma}}^{\dag} d_{i\bar{\sigma}}^{},d_{j\sigma}^{\dag}\rrangle_t \simeq 0$, corresponding to virtual charge excitations of the dots. This approach holds for strongly interacting quantum dots at temperatures above the Kondo temperature. We have introduced the notation $\llangle A,B \rrangle_t = \rmi \theta(t) \langle [A(t),B(0)]_{+}\rangle$ for brevity. After straightforward algebra we obtain the following equations for the Fourier transforms of the Green's functions and correlators
\begin{subequations}
\begin{align}
(E-\varepsilon_i)G^r_{i\sigma,j\sigma}&=\delta_{ij}+U_i\llangle d_{i\sigma}^{}n_{i\bar{\sigma}}^{},d_{i\sigma}^{\dag}\rrangle\nonumber\\ &
+\sum_{\alpha k}V_{\alpha k,i}^{*} G_{\alpha k \sigma, j\sigma}^r \ , 
\label{eq:04a}\\
(E-\varepsilon_i-U_i) & \llangle d_{i\sigma}^{} n_{i\bar{\sigma}}^{},d_{j\sigma}^{\dag} \rrangle 
=\langle n_{i\bar{\sigma}}^{}\rangle \delta_{ij} \nonumber \\
&+\sum_{\alpha k}V_{\alpha k,i}^*\llangle C_{\alpha k \sigma}n_{i\bar{\sigma}},d_{j\sigma}^{\dag}\rrangle \ ,
\label{eq:04b}
\end{align}
\label{eq:04}
\end{subequations}
where $\llangle A,B \rrangle$ denotes the Fourier transform of $\llangle A,B \rrangle_t$. Hereafter we consider the decoupling given by the Hubbard-I approximation $\llangle C_{\alpha k \sigma} n_{i\bar{\sigma}}^{},d_{i\sigma}^{\dag}\rrangle\approx \langle n_{i\bar{\sigma}}\rangle G^r_{\alpha k \sigma,i\sigma}$~\cite{Hewson42}. Equations~(\ref{eq:04}) are then readily solved, yielding a closed expression for the retarded Green's functions 
\begin{subequations}
\begin{align}
&G^{r}_{i\sigma,j\sigma}(E)=h_{ij}(E)\Bigg[\frac{1-\langle n_{i \bar{\sigma}}\rangle}{E-\varepsilon_i-\left(1+\frac{\langle n_{i \bar{\sigma}}\rangle U_i}{E-\varepsilon_i-U_i}\right)\widetilde{\Sigma}_{i}(E)}\nonumber \\
&+\frac{\langle n_{i \bar{\sigma}}\rangle}{E-\varepsilon_i-U_i-\left(1-\frac{(1-\langle n_{i \bar{\sigma}}\rangle)U_i}{E-\varepsilon_i}\right)\widetilde{\Sigma}_{i}(E)}\Bigg]\ .
\label{eq:05a}
\end{align}
For the sake of brevity we have introduced the following definitions 
\begin{align}
\Delta n_i&=1+\frac{U_i}{E-\varepsilon_i-U_i}\,\langle n_{i\bar{\sigma}}\rangle\ , \nonumber\\
\widetilde{\Sigma}_{i}(E)&=\Sigma_{ii}+\frac{\Sigma_{\bar{i}i}\Sigma_{i\bar{i}}}{E - \varepsilon_{\bar{i}}-\Delta n_{\bar{i}}\Sigma_{\bar{i}\,\bar{i}}}\,\Delta n_{\bar{i}}\ , \nonumber \\
h_{ij}(E)&=\delta_{ij}+\frac{\Delta n_{\bar{i}}\,\Sigma_{\bar{i}i}}{E-\varepsilon_{\bar{i}}-\Delta n_{\bar{i}}\,\Sigma_{\bar{i}\,\bar{i}}}\,\delta_{\bar{i}j}\ .
\label{eq:05b}
\end{align}
\label{eq:05}
\end{subequations}
Within the wide band limit the self-energy becomes $\Sigma_{ij}=-\rmi \left(\Gamma^{L}_{ij}+\Gamma^{R}_{ij}\right)/2\equiv -\rmi \Gamma_{ij}/2$. According to Eq.~(\ref{eq:05a}), the diagonal elements $G^{r}_{i\sigma,i\sigma}(E)$ have two poles that are shifted (with respect to the bare values $E=\varepsilon_i$ and $E=\varepsilon_i+U_i$) and broadened due to the presence of the complex term $\widetilde{\Sigma}_{i}$.
In the case of weak interdot capacitive interaction, the Hamiltonian 
given by Eq.~\eqref{eq:01a} acquires a term $U_{12}N_1N_2$, where 
$N_i=\sum_\sigma n_{i\sigma}$, and this term can be treated within the 
Hartree approximation. As a consequence, the two poles found above would 
become renormalized due to an additional factor $-U_{12} \langle N_{\bar{\imath}} 
\rangle$ in the denominator of Eq.~\eqref{eq:05a}. However, our results will 
qualitatively remain unaffected.

The tunnel coupling of the DQD system to both leads is encoded in the matrices~\cite{Tagani12}
\begin{eqnarray}
\Gamma^L=\Gamma_0
\left(
   \begin{array}{cc}
      1 & \sqrt{a} \\
      \sqrt{a} & a \\
   \end{array}
\right)\ , 
\quad
\Gamma^R=\Gamma_0
\left(
   \begin{array}{cc}
      a & \sqrt{a} \\
      \sqrt{a} & 1 \\
   \end{array}
\right)\ .
\label{eq:06}
\end{eqnarray}
Here $\Gamma_0$ and $a$ are parameters describing the different coupling of each dot to both leads. This coupling corresponds to the configuration studied in Ref.~\onlinecite{lad03}. For instance, $a=0$ correspond to totally decoupled quantum dots and $a=1$ to a symmetric coupling case. 
%
%
The spin-dependent occupations $\langle n_{i \bar{\sigma}}\rangle$ can be calculated from the general expression
\begin{subequations}
\begin{equation}
\langle d_{i\bar{\sigma}}^{\dag} d_{j\bar{\sigma}}^{}\rangle= \int \frac{dE}{2\pi \rmi}\, G^{<}_{j \bar{\sigma},i\bar{\sigma}}(E)\ .
\label{eq:08a}
\end{equation}
The lesser Green's function is related to the retarded and advanced Green's functions as ${\bm G}_{\sigma,\sigma}^{<}(E)=\rmi \sum_\alpha f_\alpha(E){\bm G}^{r}_{\sigma,\sigma}{\bm \Gamma}^{\alpha} {\bm G}^{a}_{\sigma,\sigma}$. Therefore
\begin{equation}
\langle n_{i \bar{\sigma}}\rangle =
 \int \frac{dE}{2\pi } \sum_{\alpha l m} f_\alpha(E)G^{r}_{i\bar{\sigma},l\bar{\sigma}}(E)\Gamma^{\alpha}_{lm} G^{a}_{m\bar{\sigma},i\bar{\sigma}}(E) \ .
\label{eq:08b}
\end{equation}
\label{eq:08}
\end{subequations}
Equations~(\ref{eq:05a}) and~(\ref{eq:08b}) are solved self-consistently to obtain $G^{r}_{i\sigma,j\sigma}(E)$.

\section{Spectral function}

The spectral function per spin is defined as
\begin{equation}
\mathcal{A}(E)=-\frac{1}{\pi}\,\Im{\Tr{\mathbf{G}^{r}_{\sigma\sigma}(E)}}\ .
\label{eq:09}
\end{equation}
The absence of magnetic interactions ensures that $\mathcal{A}(E)$ becomes spin-independent. In this section we discuss the spectral function in equilibrium at $T=0$ for concreteness ($\mu_L=\mu_R=0$ and $T_L=T_R=0$).

To gain insight into the occurrence of BICs in the DQD system, first we consider the non-interacting case by setting $U_i=0$ for the moment. We take $\varepsilon_1=-\varepsilon_2\equiv \varepsilon$ to obtain simpler expressions, although more general situations can be handled in the same way. A lengthy but straightforward calculation yields the following expression for the spectral density in the non-interacting case
\begin{subequations}
\begin{eqnarray}
\mathcal{A}(E)&=&\frac{(1+a)\Gamma_0}{\pi D(E)}\left[E^2+\varepsilon^2+\frac{1}{4}(1-a)^2\Gamma_0^2\right]\ , \label{eq:10a}\\
D(E)&\equiv&(E^2-\varepsilon^2)^2+\left[\frac{\Gamma_0}{2}(1-a)\right]^4\nonumber \\
&+&\frac{\Gamma_0^2}{2}\Big[E^2(1+6a+a^2)+\varepsilon^2(1-a)^2\Big]\ .
\label{eq:10b}
\end{eqnarray}
Taking $a=1$ (symmetric coupling to leads) and $|E|<\varepsilon<\Gamma_0$, the spectral function is approximately given as
\begin{equation}
\mathcal{A}(E)\simeq \frac{1}{\pi}\,\frac{2\Gamma_0}{E^2+4\Gamma_0^2}+
\frac{1}{\pi}\,\frac{\varepsilon^2/2\Gamma_0}{E^2+(\varepsilon^2/2\Gamma_0)^2}\ .
\label{eq:10c}
\end{equation}
The spectral function is the sum of two Lorentzians centered at $E=0$, originated from the superposition of two states. One of these states is strongly coupled to the continuum, giving rise a wide peak of width $2\Gamma_0$. However, the level broadening of the other state is small when $\varepsilon<\Gamma_0$, indicating that it is only weakly coupled to the continuum. Tuning the levels of both dots at resonance ($\varepsilon\to 0$), the spectral function becomes
\begin{equation}
\mathcal{A}(E)\simeq \frac{1}{\pi}\,\frac{2\Gamma_0}{E^2+4\Gamma_0^2}+\delta(E)\ .
\label{eq:10d}
\end{equation}
The presence of a $\delta$-function in $\mathcal{A}(E)$ characterizes a truly bound state at $E=0$. This localized state becomes effectively decoupled from the continuum states when $a\to 1$ and $\varepsilon_1 \to \varepsilon_2$ but its energy lies at the band center. Therefore, the DQD system with negligible Coulomb interaction supports a BIC~\cite{Gonzalez-Santander13,Alvarez15}. It should be stressed that non-symmetric tunnel coupling to the leads ($a\neq 1$) smears out the BIC and only a single peak of finite width arises when $\varepsilon=0$. From  Eq. (\ref{eq:10a}) one finds in this case
\begin{equation}
\mathcal{A}(E)\sim \frac{1}{\pi}\,\frac{\gamma(a)}{E^2+\gamma^2(a)}\ ,
\label{eq:10e}
\end{equation}
\label{eq:10}
\end{subequations}
if $|E| < \Gamma_0$~\cite{Alvarez15}. The width of the Lorentzian profile $\gamma(a)\equiv \Gamma_0 (1-a)^2/\sqrt{8\left(1+6a+a^2\right)}$ only vanishes in the limit $a\to 1$, as expected.

\begin{figure}[t]
\begin{center}
\includegraphics[width=0.8\linewidth]{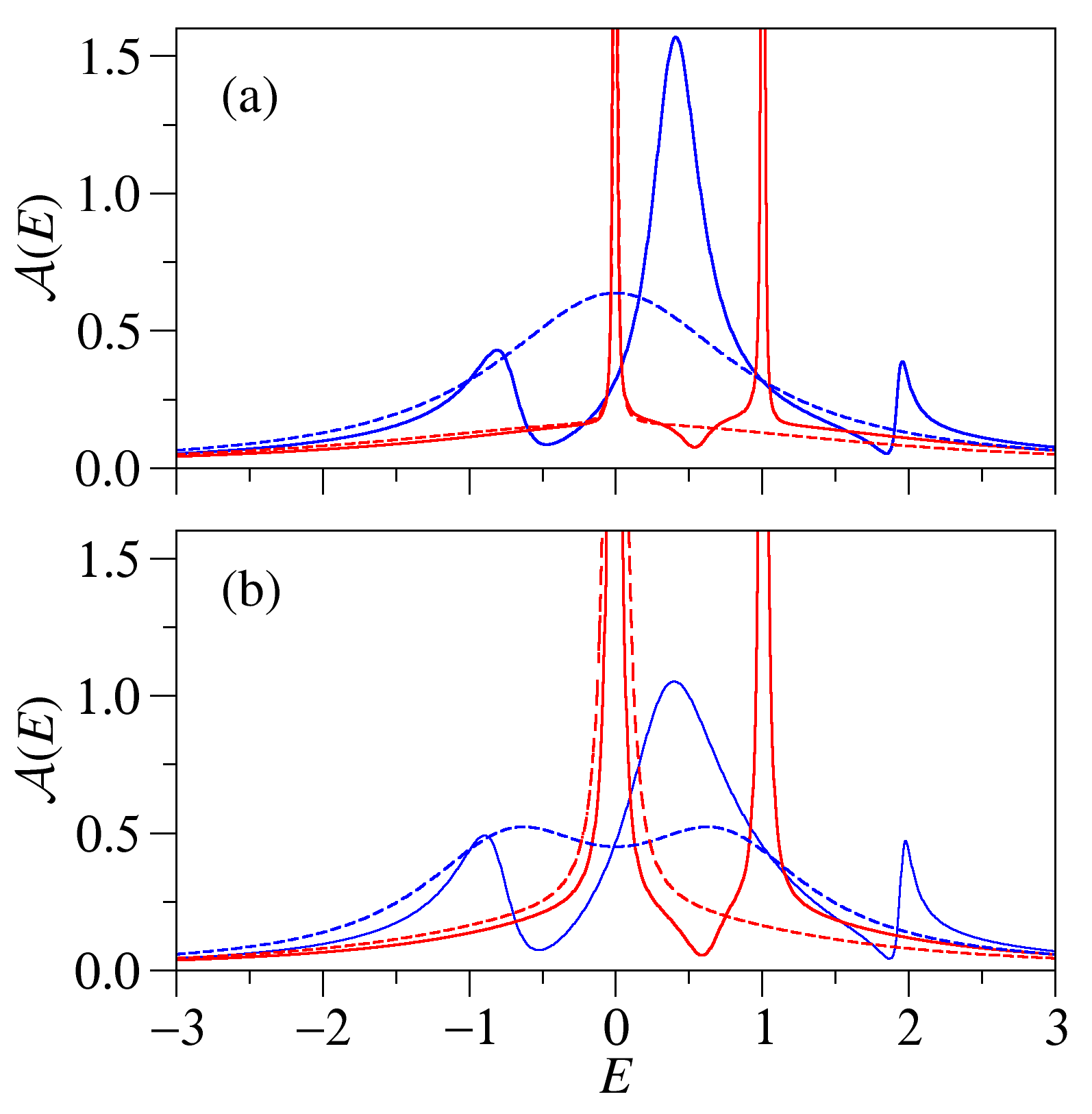}
\end{center}
\caption{(Color online) Spectral function at $T=0$ for a DQD system tunnel coupled to leads (a)~symmetrically ($a=1$) and (b)~non-symmetrically ($a=0.5$). Solid and dashed lines correspond to interacting ($U=1$) and non-interacting ($U = 0$) electrons, respectively, meanwhile the color line is blue ($\varepsilon=1.0$) or red ($\varepsilon=0.05$). Energies are expressed in units of $\Gamma_0$.}
\label{fig2}
\end{figure}

We now turn to one of our main goals, namely, the effects of the Coulomb 
interaction on the BICs discussed above. At finite values of the Coulomb 
energy $U_i$, Eqs~(\ref{eq:05a}) and~(\ref{eq:08b}) have to be solved 
numerically. Figure~\ref{fig2} shows  in solid line the spectral function 
$\mathcal{A}(E)$ at $T=0$ when $U_1=U_2 \equiv U = 1$. Hereafter, we take 
$\Gamma_0=1$ as our unit of energy. Results for the non-interacting DQD 
system ($U=0$) are shown in dashed line for comparison. According to 
Eq.~(\ref{eq:10b}), the spectral function of the noninteracting DQD system 
($U=0$) is a superposition of two Lorentzians of finite width when the tunnel 
coupling to the leads is symmetric~($a=1$) and the dots are 
detuned~($\varepsilon=1$), as seen in blue in Fig.~\ref{fig2}(a). The two 
Lorentzians cannot be resolved since their widths $2\Gamma_0$ and 
$\varepsilon^2/2\Gamma_0$ do not differ much with the chosen parameters. 
After the Coulomb interaction is switched on, virtual levels around 
$\varepsilon_i+U$ arise, giving rise to a complex pattern with maxima and 
minima. Maxima and inflection points are located at energies $\pm\varepsilon$ 
and $\pm\varepsilon+U$. 

More interesting results are found in the limit $\varepsilon\to 0$, shown in 
red in Fig.~\ref{fig2}(a). In the non-interacting DQD system the spectral 
function presents a narrow peak at $E=0$ when $\varepsilon$ is small but 
finite. This central peak resembles the $\delta$-function term in 
Eq~(\ref{eq:10c}) when $\varepsilon=0$ and signals the occurrence of a BIC. 
Most importantly, the peak is replicated at an energy $E=U$ when Coulomb 
interaction is finite. 
The numerical evidence that the BIC is still present and replicated 
at $E=U$ can be substantiated by noticing that the occupation of both quantum 
dots is the same $\langle n_{1\bar{\sigma}}\rangle = \langle 
n_{2\bar{\sigma}}\rangle \equiv \langle n\rangle$ when $\varepsilon=0$ and 
the coupling to the leads is symmetric ($a=1$). Proceeding in the same way as 
in Eq.~(\ref{eq:10d}), the spectral function given by Eq.~(\ref{eq:09}) can be calculated 
from the retarded Green's function~(\ref{eq:05a}) in this limit case. 
Neglecting the contribution of the resonant states strongly coupled to the 
continuum, Eq.~(\ref{eq:05a}) has two poles at energies around the resonances
($|E|,|E-U|\ll \Gamma_0$), weighted by the occupations. We thus find
for the spectral function
\begin{equation}
\mathcal{A}(E) \simeq \big[1-\langle n\rangle\big]\delta(E)+
\langle n\rangle\,\delta(E-U)\, .
\label{BIC_replicated}
\end{equation}
in the limit $\varepsilon=0$, in perfect agreement with 
the numerics. As expected, the BIC replicated at $E=U$ vanishes when the occupation is 
negligible.

Hence, we come to the important conclusion that BICs are preserved even in 
the presence of the Coulomb interaction. 

Figure~\ref{fig2}(b) shows the spectral density in the 
case of asymmetric tunnel coupling to the leads ($a=0.5$). When $U=0$ and 
$\varepsilon$ is large (detuned quantum dots), the spectral function is bimodal, 
according to the general expression~(\ref{eq:10a}). Similarly to the case of 
symmetric coupling, more features arise when $U$ is finite. The two maxima 
merge into a single peak at $E=0$ when the levels of the dots approach each 
other ($\varepsilon\to 0$) and this peak is replicated at $E=U$ when the 
Coulomb interaction is finite.

\section{Transmission coefficient}

\begin{figure}[t]
\begin{center}
\includegraphics[width=0.8\linewidth]{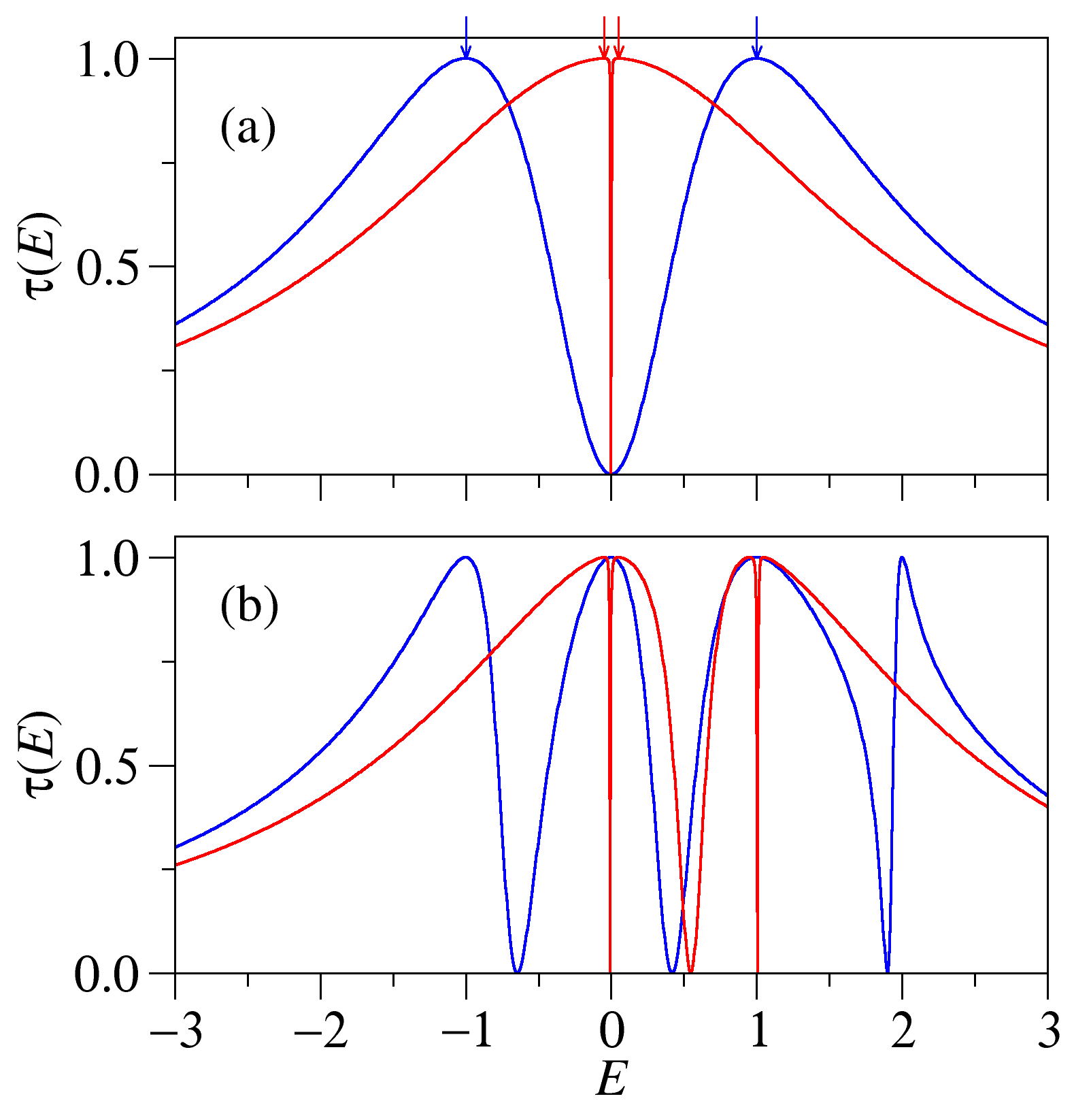}
\end{center}
\caption{(Color online) Transmission coefficient at $T=0$ for a DQD system tunnel coupled symmetrically to leads ($a=1$), corresponding to (a)~non-interacting ($U=0$) and (b)~interacting ($U = 1$) electrons, respectively. Color line is blue ($\varepsilon=1.0$) or red ($\varepsilon=0.05$). Blue and red arrows in the upper panel indicate $E=\pm \varepsilon$. Energies are expressed in units of $\Gamma_0$.}
\label{fig3}
\end{figure}

The EOM approach introduced in Sec.~\ref{mfa} turns the initial many body problem into an effective one body problem. In this section we focus on the transmission function, Eq.~(\ref{eq:11}), which enters the current formula, Eq. ~(\ref{eq:02b}). Under this premise, the transmission coefficient can be expressed as~\cite{Meir92}
\begin{equation}
\tau(E)=\Tr\left[{\bm G}^{a}_{\sigma\sigma}(E){\bm \Gamma}^{R}{\bm G}^{r}_{\sigma\sigma}(E){\bm \Gamma}^{L}\right]\ .
\label{eq:11}
\end{equation}
We restrict ourselves to equilibrium by setting $\mu_L=\mu_R=0$ and $T_L=T_R=0$ throughout this section. In the absence of Coulomb interaction the transmission coefficient can be obtained analytically: $\tau(E)=4a\Gamma_0^2 E^2/D(E)$~\cite{Alvarez15} with $D(E)$ given by Eq.~(\ref{eq:10b}). When $D(E=0)\neq 0$ the transmission coefficient vanishes at $E=0$. For instance, when the coupling to the leads is symmetric ($a=1$) but the energy levels of the dots are detuned ($\varepsilon\neq 0$) one obtains 
\begin{subequations}
\begin{equation}
\tau(E) \simeq \frac{E^2}{E^2+\varepsilon^4/4\Gamma_0^2}\ ,
\label{eq:12a}
\end{equation}
for $|E|<\varepsilon$. The transmission coefficient becomes unity at energies $\pm\varepsilon$, as indicated by vertical arrows in Fig.~\ref{fig3}(a). We remark that the transmission coefficient displays a Fano antiresonance profile of width $\varepsilon^2/2\Gamma_0$, 
originated from the interference of two coexisting paths for a traveling electron
in the  system. One path is a direct way that traverses the DQD via the 
strongly coupled state while the second path includes a hopping on and off the BIC and then the electron continues with 
propagation. The destructive interference between these two paths is at the heart of
the Fano antiresonance as shown in Eq.~(\ref{eq:12a})~\cite{bulka}.

Similarly, when the energy levels are aligned ($\varepsilon=0$) but the 
coupling to the leads is asymmetric ($a\neq 1$), the transmission shows again 
a Fano antiresonance around $E=0$, namely $\tau(E) \sim 
E^2/\left[E^2+\gamma^2(a)\right]$. Although transmission may take large 
values at some specific energies if $a\neq 1$, it should be mentioned that 
the DQD system never reaches perfect transparency.

From Eq.~(\ref{eq:12a}) it becomes apparent that the dip gets narrower when the levels of the quantum dots approach each other, as seen in Fig.~\ref{fig3}(a) for $\varepsilon=0.05$ (red curve), and eventually disappears if $\varepsilon = 0$. In this limit case, corresponding to $D(E=0)\to 0$ when $a=1$, the transmission coefficient displays a Lorentzian shape of width $2\Gamma_0$,
\begin{equation}
\tau(E)=\frac{4\Gamma_0^2}{E^2+4\Gamma_0^2}\ .
\label{eq:12b}
\end{equation}
\label{eq:12}
\end{subequations}
Therefore, unless the system is finely tuned ($\varepsilon=0$ and $a=1$), in general the transmission coefficient shows a marked dip at $E=0$.

For the interacting DQD system the transmission coefficient given by 
Eq.~(\ref{eq:11}) has to be computed numerically from Eqs.~(\ref{eq:05a}) 
and~(\ref{eq:08b}). In Fig.~\ref{fig3}(b) we observe that the transmission 
coefficient at $E=0$ becomes unity when $\varepsilon=1$ (blue curve), whereas 
it vanishes if $U=0$ [see Fig.~\ref{fig3}(a)]. Coulomb interaction splits the 
levels and the transmission coefficient then displays four maxima at energies 
$E=\pm\varepsilon$ and $E=\pm\varepsilon +U$, as seen in Fig.~\ref{fig3}(b). 
Moreover, in the limit $\varepsilon \to 0$, the narrow antiresonance at $E=0$ 
found when $U=0$ is replicated at $E=U$ in the interacting DQD system. This 
result is consistent with the spectral function discussed in the previous 
section and confirms the robustness of BICs under Coulomb interaction.

\section{Voltage-driven electric transport}

\begin{figure}[t]
\begin{center}
\includegraphics[width=0.9\linewidth]{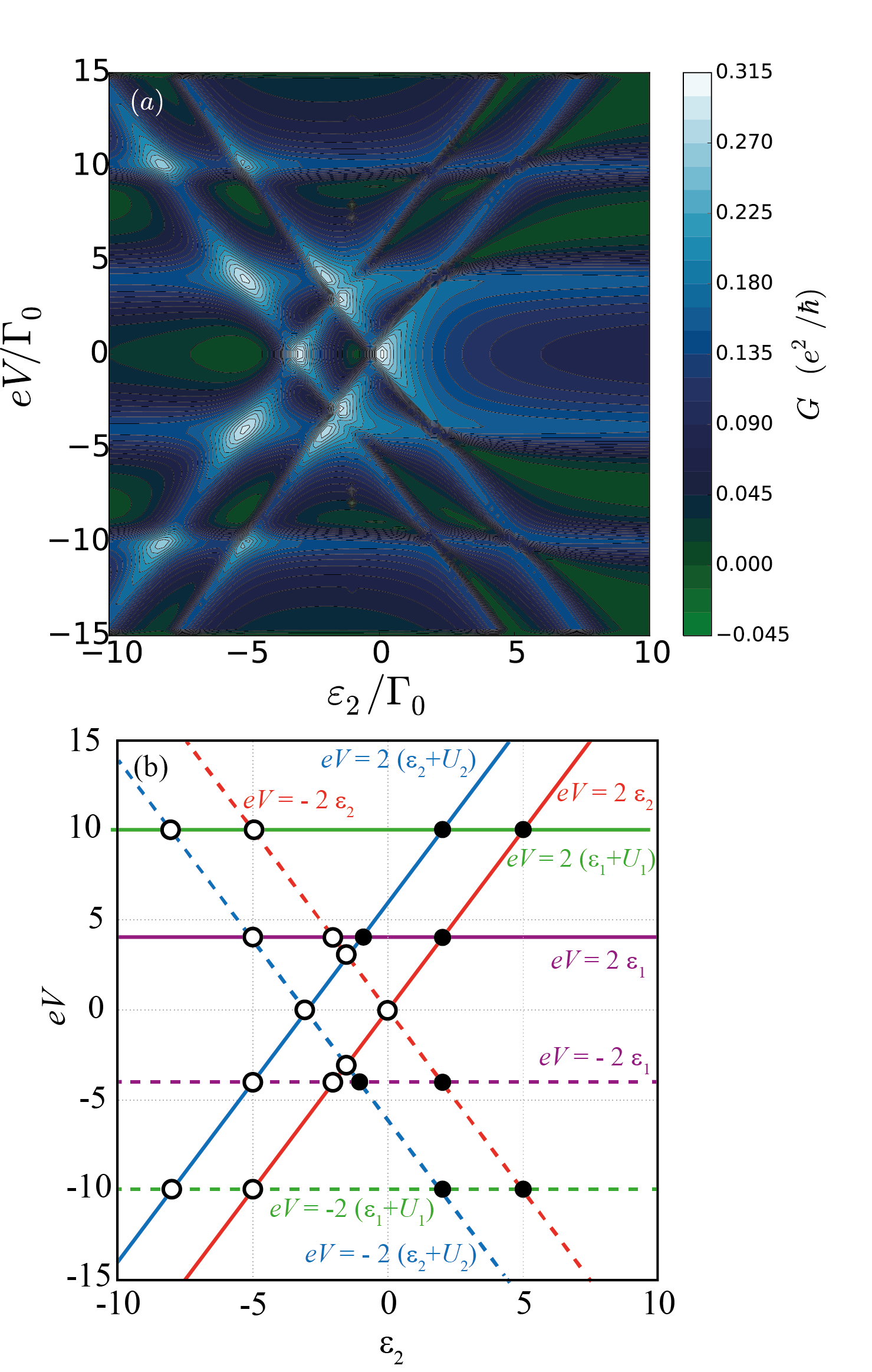}
\end{center}
\caption{(Color online) Top: Differential electric conductance as a function of $eV$ and $\varepsilon_2$ for $\varepsilon_1=2$, $U_1=U_2=3$ and $T_L=T_R=T=10^{-3}$. Bottom: Diagram of the energy levels showing crossing points that correspond to maxima (open circles) and minima (solid circles) of the differential conductance. Energies are expressed in units of $\Gamma_0$.}
\label{fig4}
\end{figure}

We now discuss the impact of BICs on the electric response of the DQD system 
out of equilibrium. To this end, we calculate the non-linear dependence of 
the electric current given by Eq.~(\ref{eq:02b}) on the source-drain voltage 
$V$. Crossing of the four levels involved, namely $\varepsilon_i$ and 
$\varepsilon_i+U_i$, makes the Fano antiresonance in the transmission 
narrower and eventually it disappears when the system is finely tuned, as 
discussed in the previous section. The crossing reveals itself as minima of 
the differential conductance curves $G=dI/dV$, as seen in Fig.~\ref{fig4} (we 
here set the Fermi energy $\varepsilon_F=0$ and the rest of the energies in 
units of $\Gamma_0$). Figure~\ref{fig4}(a) displays the low-temperature 
differential conductance as a function of $eV$ and $\varepsilon_2$ for 
$\varepsilon_1=2$, $U_1=U_2=3$ and $T_L=T_R=T=10^{-3}$. We observe that the 
differential conductance displays abrupt minima when the two levels cross 
simultaneously $eV/2$ or $-eV/2$. This condition is better seen in 
Fig.~\ref{fig4}(b), where we plot the eight straight lines corresponding to 
the resonant conditions $eV=\pm 2 \varepsilon_i$ and $eV= \pm 2 
(\varepsilon_i+U_i)$. It is worth noting that similar conductance 
patterns are found for interacting two-orbital quantum dots~\cite{ger15}. 
Solid circles indicate the crossing points where the differential conductance 
reaches a local minimum, in perfect agreement with Fig.~\ref{fig4}(a). 
Therefore, the existence of BICs can be detected  in transport experiments by 
measuring the differential conductance. It should be mentioned that the 
differential conductance displays local maxima when one level crosses $eV/2$ 
and the other reaches the value $-eV/2$. These maxima are marked with open 
circles in Fig.~\ref{fig4}(b).

\section{Temperature-driven electric transport}

We now investigate the electric transport in response to a thermal gradient $\theta$. For $\theta>0$ we take $T_L=T+\theta$
and $T_R=T$ with $T$ the background temperature while for $\theta<0$ we take $T_L=T$
and $T_R=T+|\theta|$ . Thus, $\theta$ is positive (negative) when the left (right) electrode is the hotter contact. In this section, the bias voltage is absent ($\mu_L=\mu_R=\varepsilon_F$ or $V=0$) and charge flows only due to the external thermal gradient. The differential thermoelectric conductance $L=dI/d\theta$ for $\theta>0$ is expressed as
\begin{subequations}
\begin{equation}
L=\frac{e}{h} \int_{-\infty}^\infty dE\,\left[g(E,\theta) + (f_L-f_R)
\frac{\partial\phantom{\theta}}{\partial \theta}\right]\tau(E,\theta) \, ,
\label{eq:13a}
\end{equation}
with
\begin{equation}
g(E,\theta)  \equiv \frac{E-\varepsilon_F}{T_L}
\left[-\frac{\partial f_L(E,\theta)}{\partial E}\right]\, , \label{eq:13b}
\end{equation}
\label{eq:13}
\end{subequations}
and similarly for $\theta<0$ after making the substitutions $f_L(E,\theta)\leftrightarrow-f_R(E,\theta)$ and $T_L\leftrightarrow T_R$ in the first term of 
the integrand.
Since $g(E,\theta)$ is an odd function of $E-\mu_L$, the thermoelectric conductance vanishes if the transmission coefficient is a symmetric function of the energy $E$ with respect to the Fermi level $\varepsilon_F$. 

\begin{figure}[t!]
\begin{center}
\includegraphics[width=1.05\linewidth]{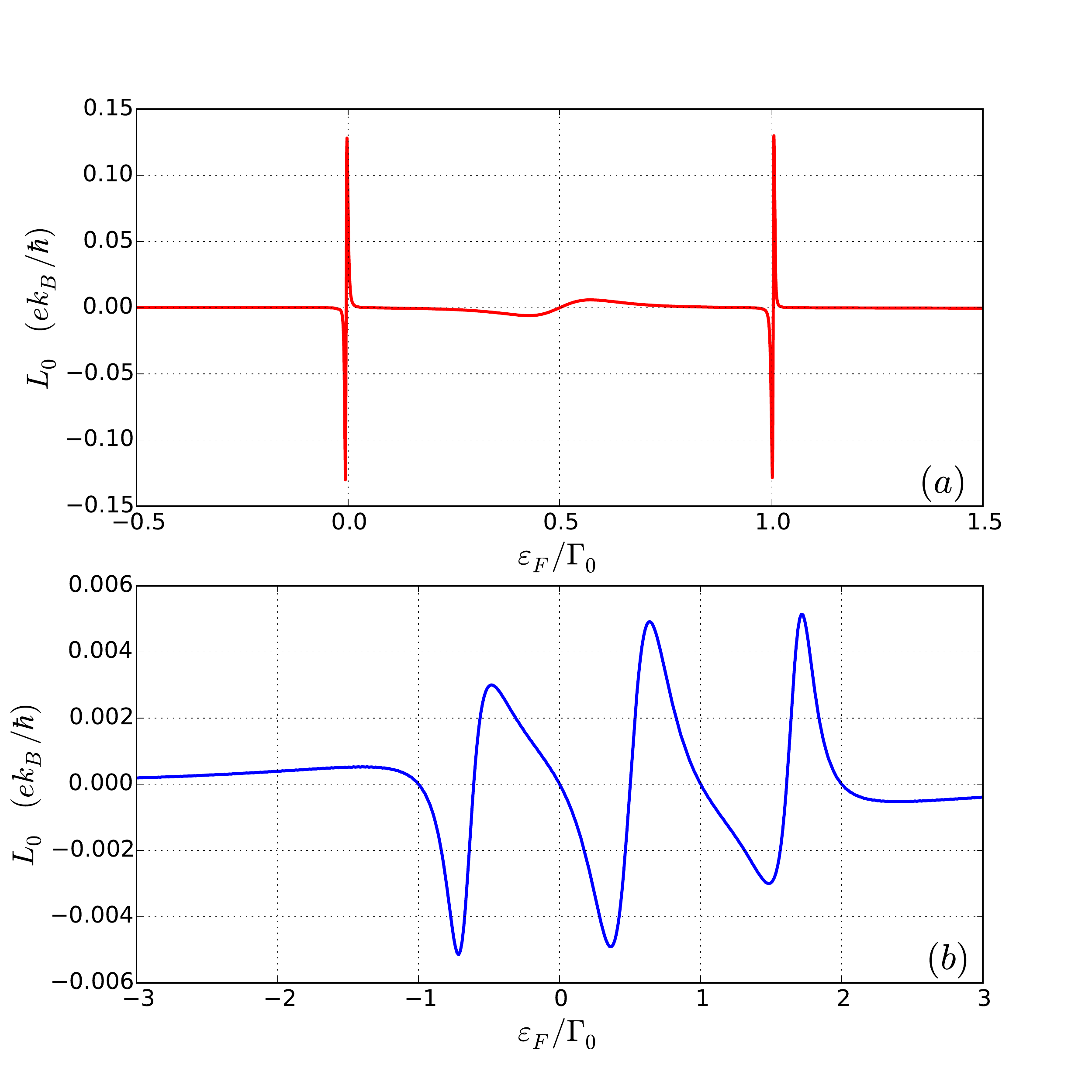}
\end{center}
\caption{(Color online) Linear thermoconductance as a function of the temperature bias $k_B \theta$ for (a) $\varepsilon=0.05$
and (b) $\varepsilon=1$. The charging energies are $U_1=U_2=1$ and the background temperature is $k_BT=0.001$. Energies are expressed in units of $\Gamma_0$.}
\label{fig:L0}
\end{figure}

In Fig.~\ref{fig:L0} we depict the linear thermoelectric conductance $L_0$ of 
the DQD system as a function of the Fermi energy. $L_0$ is obtained from 
Eq.~\eqref{eq:13a} in the limit $\theta\to 0$. As a consequence, the second 
term in the right-hand side of Eq.~\eqref{eq:13a} cancels out and the 
functions $g$ and $\tau$ are evaluated at $\theta=0$. At low temperature, a 
Sommerfeld expansion shows that $L_0$ is proportional to the derivative of 
$\tau$ with respect to energy at $E=\varepsilon_F$~\cite{mott}. Therefore, 
$L_0$ is a magnitude sensitive to changes in the transmission, which makes it 
useful in the detection of narrow resonances.

Figure~\ref{fig:L0}(a) shows the case $\varepsilon=0.05$. We observe two 
asymmetric resonances when $\varepsilon_F=0$ and $\varepsilon_F=1$ and a 
smooth variation around $\varepsilon_F=0.5$. We attribute the former to the 
transmission dips seen in Fig.~\ref{fig3}(b) and the latter to the more 
broadened antiresonance at $E=0.5$. However, one would expect that $L_0$ 
crosses zero whenever $\varepsilon_F$ aligns with an extremum of $\tau$ 
(maximum or minimum). The contribution of the transmission maxima cannot be 
observed in Fig.~\ref{fig:L0}(a) due to the sharp variation around the dips. 
This can be demonstrated with a larger value of $\varepsilon$ 
[$\varepsilon=1$ in Fig.~\ref{fig:L0}(b)]. Both the maxima and the minima in 
$\tau$ have now comparable widths [see Fig.~\ref{fig3}(b)] and their effect 
in $L_0$ are hence visible: There exist seven zeros in Fig.~\ref{fig:L0}(b), 
which exactly correspond to the extrema of $\tau$. Importantly, $L_0$ reaches 
values as significant as $0.12$ (in units of $ek_B/\hbar=20$~nA/K) for the case 
of the narrow antiresonances ($\varepsilon=0.05$), showing that BICs can be 
useful in thermoelectric applications that convert waste heat into 
electricity.

\begin{figure}[t!]
\begin{center}
\includegraphics[width=0.9\linewidth]{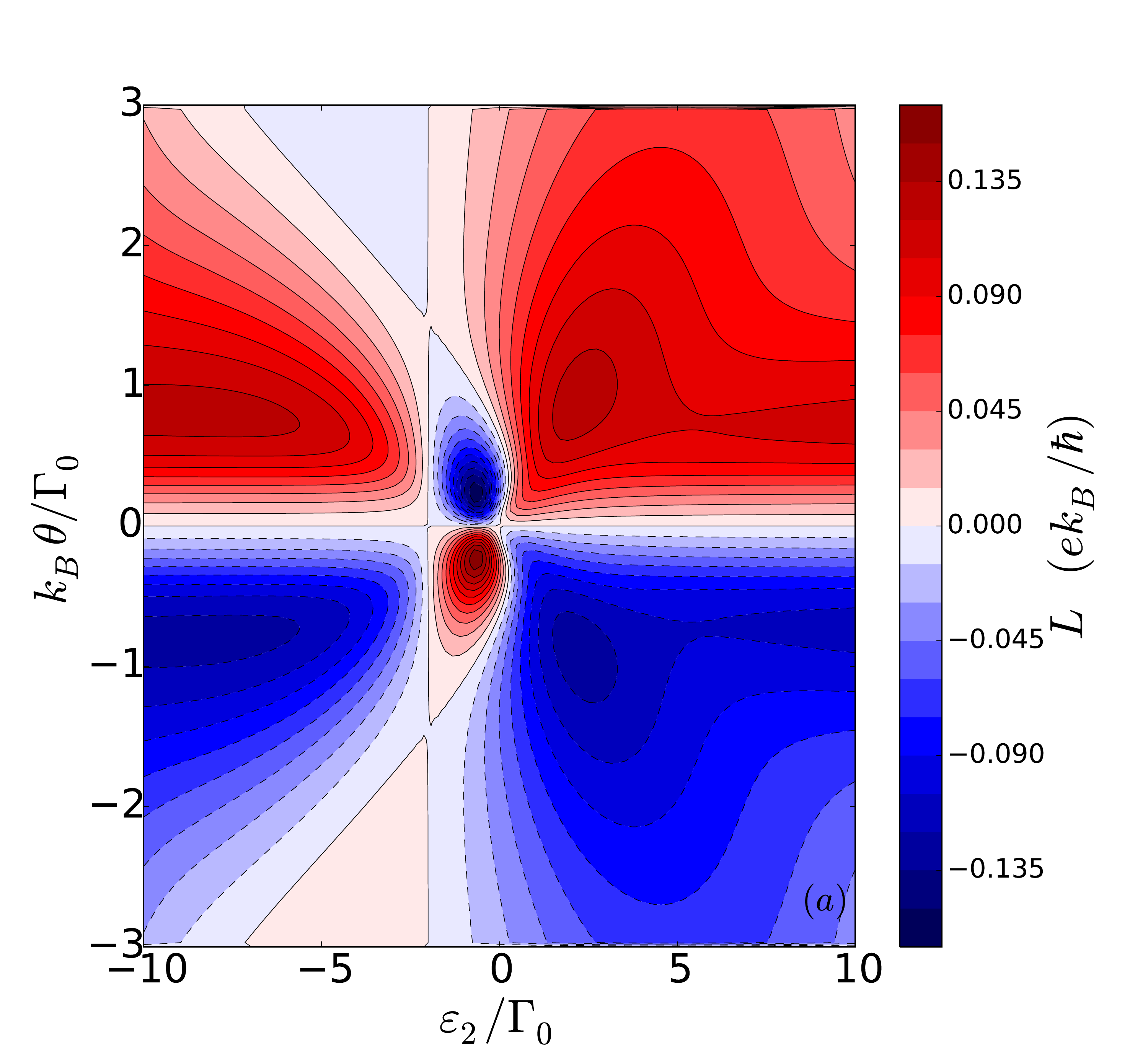}
\includegraphics[width=0.9\linewidth]{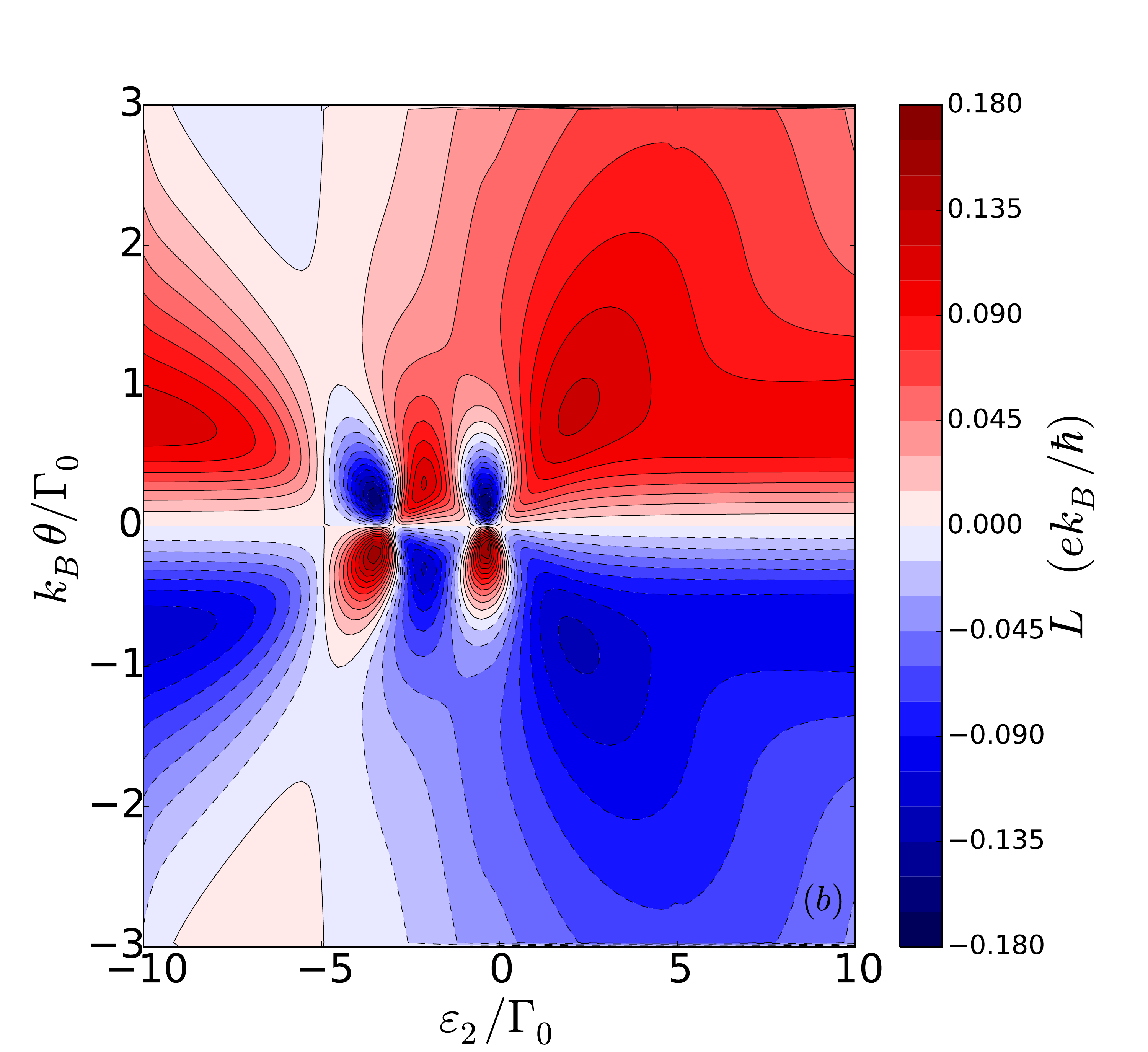}
\end{center}
\caption{(Color online) Differential thermoelectric conductance as a function of the temperature bias $k_B \theta$ and the level position $\varepsilon_2$ for $\varepsilon_1=2$. (a) $U_1=U_2=0$ (noninteracting case). (b) $U_1=U_2=3$ (interacting case). The background temperature is $k_BT=0.001$. Energies are expressed in units of $\Gamma_0$.}
\label{fig:Ldif}
\end{figure}

Let us turn to the properties of the differential conductance given by 
Eq.~\eqref{eq:13a}. In Ref.~\onlinecite{sie14} two of us predicted the 
appearance of a butterfly structure in $L$ as a function of the applied 
temperature bias and the energy level of a single dot. This characteristic 
shape can ultimately be traced back to the presence of two resonances in the 
dot due to Coulomb repulsion, which has been confirmed 
experimentally~\cite{svi15}. In the case of the parallel DQD system 
considered here, we have quantum interference effects that distort the 
butterfly structure. In Fig.~\ref{fig:Ldif} we show $L$ as a function of 
$\theta$ and $\varepsilon_2$ for a fixed value $\varepsilon_1=2\Gamma_0$. 
When $U_1=U_2=0$ we find a vanishing thermoelectric conductance for 
$\varepsilon_2=-2\Gamma_0$ independently of the temperature bias [see 
Fig.~\ref{fig:Ldif}(a)]. We recall that the configuration 
$\varepsilon_1=-\varepsilon_2$ leads to transmission functions symmetric 
relative to the Fermi energy [see Fig.~\ref{fig3}(a)]. As a consequence, 
$L=0$ to all orders in $\theta$. Away from $\varepsilon_2=-2\Gamma_0$ we can 
explain the behavior for small $\theta$ as follows. As $\varepsilon_2$ 
increases, the transmission function becomes asymmetric with two peaks in 
$\varepsilon_1$ and $\varepsilon_2$. Since $\varepsilon_2$ is closer to 
$\varepsilon_F$ the thermocurrent will flow from the right electrode when the 
left electrode is hotter. Therefore, $L$ is negative as shown in the blue 
central area of Fig.~\ref{fig:Ldif}(a). Further increase of $\varepsilon_2$ 
favors the transport of thermally excited electrons above the Fermi energy 
and $L$ turns out to be positive [the red region to the right of 
Fig.~\ref{fig:Ldif}(a)]. For negative $\theta$ the thermoelectric conductance 
changes sign because the right contact is now hotter. The interacting case is 
plotted in Fig.~\ref{fig:Ldif}(b). Based on the results depicted in 
Fig.~\ref{fig3}(b) the transmission comprises multiple peaks that give rise 
to additional changes of sign for $L$ [compare Fig.~\ref{fig:Ldif}(b) 
and~(a)]. For example, $L$ is almost zero not only for 
$\varepsilon_2=-2\Gamma_0$ but also now for $\varepsilon_2=-5\Gamma_0$. 
Nevertheless, we should remark that in the interacting situation the symmetry 
of the transmission lineshape is only approximate unlike the noninteracting 
case. As a consequence, the thermoelectric conductance quickly becomes 
nonzero when $\theta$ departs from the linear regime.

 \begin{figure}[t!]
\begin{center}
\includegraphics[width=1.05\linewidth]{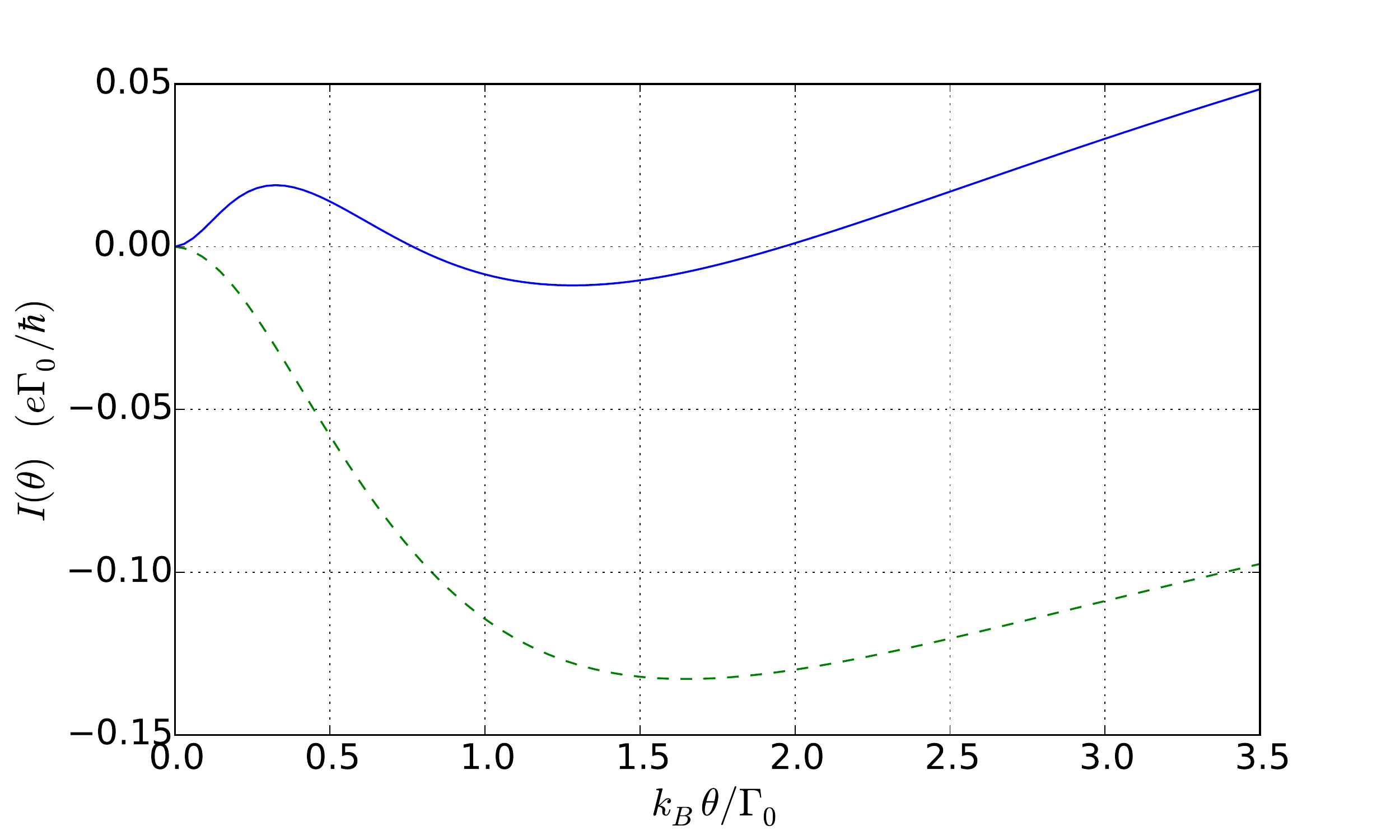}
\end{center}
\caption{(Color online) Thermocurrent as a function of the temperature bias $k_B \theta$ for different values of the charging energy. Blue curve: $U_1=U_2=1.2$; green curve: $U_1=U_2=0$. The energy level positions of the dot are $\varepsilon_1=4.0$ and $\varepsilon_2=-1.05$. The background temperature is $k_BT=0.001$. Energies are expressed in units of $\Gamma_0$.}
\label{fig:Ith}
\end{figure}

The rich structure of the differential thermoelectric conductance discussed 
above suggests that the thermocurrent $I(\theta)$ might cross the $\theta$ 
axis several times unlike the single dot case, where there exists one 
nontrivial zero only~\cite{san16,fah13,svi15,sie14,aze14,hwa14,zim15,sta14,jia15}. 
In Fig.~\ref{fig:Ith} we show a representative result of $I(\theta)$ for 
different values of the charging energy. Remarkably, the thermocurrent 
becomes strongly nonlinear and changes sign twice for a tuning of the dot 
levels such that $\varepsilon_1=4$ and $\varepsilon_2=-1.05$ (blue curve). 
For comparison, we also show the noninteracting case (green curve) for which 
the sign of $I$ stays the same. Strikingly enough, the current--temperature 
curve when interactions are present presents a marked region of negative 
differential conductance reminiscent of the current--voltage characteristics 
of Esaki diodes~\cite{esaki}. The difference is that in our case transport is 
driven purely by thermal means. We obtain a peak-to-valley ratio (defined 
as the distance between the maximum and the minimum magnitudes divided by the 
current peak) of the order of $1.7$. 

\begin{figure}[t]
\begin{center}
\includegraphics[width=1.025\linewidth,clip]{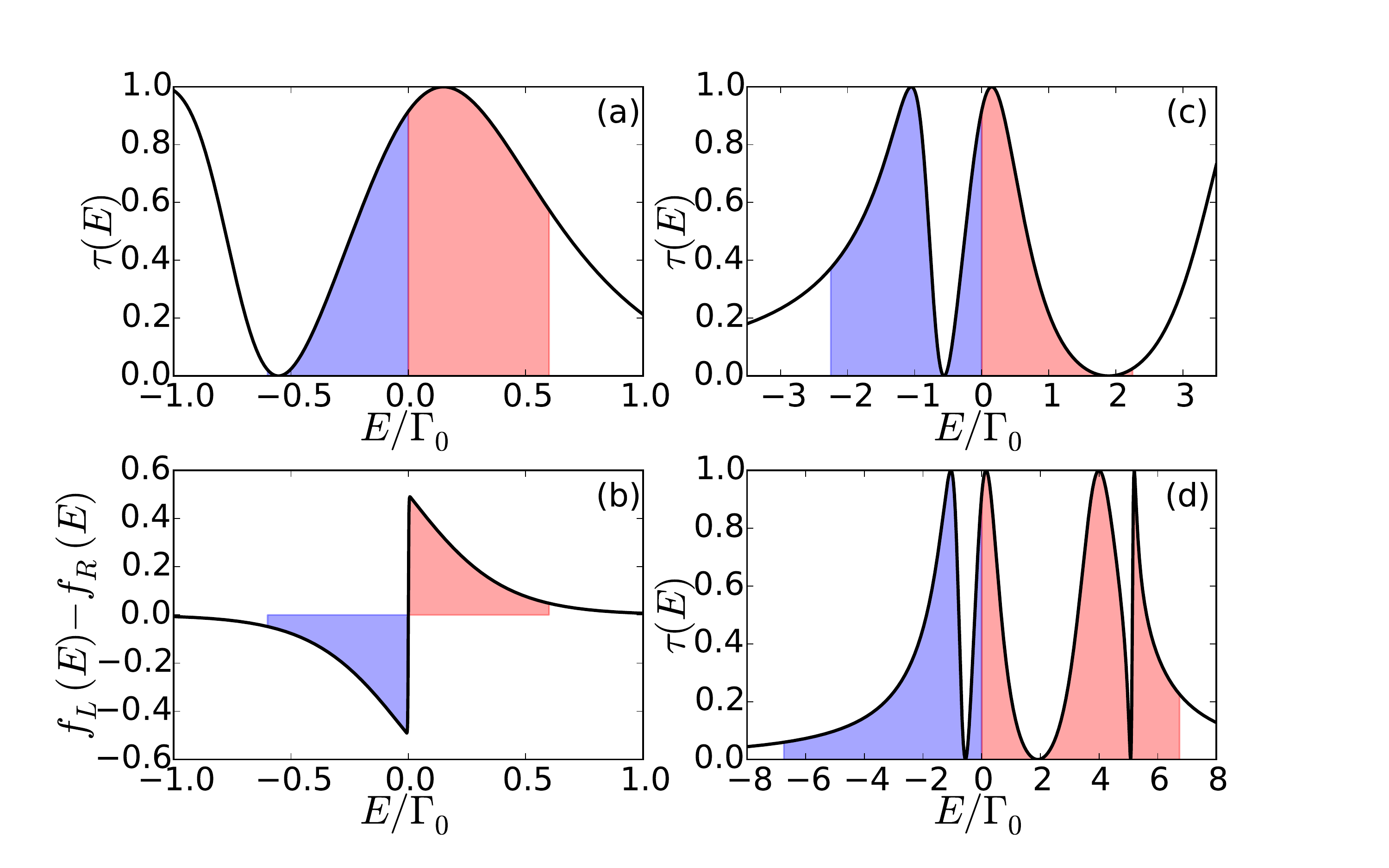}
\end{center}
\caption{(Color online) Sketch explaining the nonlinear behaviour of the thermocurrent. Transmission function as a function of  energy for (a) $\theta=0.2$, (c) $\theta=0.75$ and (d) $\theta=2.25$. Charging energies are $U_1=U_2=1.2$ and background temperature is  $k_B T=0.0001$ (in units of $\Gamma_0$). The energy level positions of the dot are $\varepsilon_1=4.0$ and $\varepsilon_2=-1.05$. The color filling represents the part of the transmission function which contributes in the transport at given $\theta$. These parts are calculated from the overlap with the Fermi function difference [panel (b) for $\theta=0.2$] which is nonzero within the energy range indicated with vertical dashed lines. Blue (red) filling represents transport of electrons (holes).}
\label{fig:LEsq}
\end{figure}

Since the existence of negative differential conductance driven by a
temperature gradient is an important result, we explain  the origin of the 
current behavior in the sketch of Fig.~\ref{fig:LEsq}. In 
Fig.~\ref{fig:LEsq}(a) we plot the transmission probability as a function of 
energy calculated for the nonzero value of a small thermal bias. This curve 
has to be convoluted with the Fermi function difference [see 
Fig.~\ref{fig:LEsq}(b)] in order to calculate the full thermoelectric current 
$I$ given by Eqs.~\eqref{eq:02a} and~\eqref{eq:11}. For a given $\theta$, 
$f_L-f_R$ is an antisymmetric function of $E$ which is nonzero for energies 
smaller or of the order of $\pm 3 k_B\theta$. Note that the difference is 
positive (negative) for energies above (below) the Fermi energy 
$\varepsilon_F=0$, as should be. When multiplied by the transmission, this 
means that for $\theta=0.20$ the contribution from the electrons (carriers 
traveling above $\varepsilon_F$) is larger (red color) than that from the 
holes (carriers traveling below $\varepsilon_F$, blue color). As a 
consequence, the net current is positive (see the blue curve of 
Fig.~\ref{fig:Ith} for $\theta=0.20$, i.e., before the first nontrivial 
zero). When the temperature bias increases, the transmission is modified 
accordingly [see Fig.~\ref{fig:LEsq}(c) for $\theta=0.75$]. In this case, the 
hole-type transport surpasses the flow of electrons because the energy range 
for $f_L-f_R$ has to be extended (the occurrence of an antiresonance in 
$\tau$ also plays a role). Therefore, $I$ becomes negative and a nontrivial 
zero emerges at around $\theta=0.7$. Further enhancement of $\theta$ involves 
an even larger energy interval, which now covers additional transmission 
peaks at higher energies [see Fig.~\ref{fig:LEsq}(d) for $\theta=2.25$]. 
Then, the electron current again becomes dominant and $I$ attains a positive 
value, which implies a second nontrivial zero at around $\theta=2.0$ in 
Fig.~\ref{fig:Ith}.

\section{Conclusions}

In summary, we have analyzed the impact of electron-electron interactions in 
the spectral density of a parallel DQD system, highlighting the role of BICs. 
We find that the existence of these states is a robust phenomenon, offering the oportunity of exploiting it in electric and thermoelectric nanodevices. 
To show this, we have discussed the changes in the Coulomb diamond structure 
of the differential conductance and the presence of maxima and minima in the 
energy diagram that can be detected in a transport experiment. More 
importantly, the thermoelectric conductance exhibits a series of asymmetric 
peaks when the Fermi energy is varied across the system resonances. Maps of 
the thermoelectric response show a distorted butterfly structure due to 
intradot charging effects. In the nonlinear regime of transport, we have 
found nontrivial zeros and negative differential conductances as a function 
of the applied thermal gradient, which opens up the possibility of employing 
parallel double dots as nanoscale amplifiers and generators of electronic 
oscillations based on thermal gradients only.

\acknowledgments

We thank R.~L\'opez for useful discussions. This work has been supported by 
MINECO under Grants FIS2014-52564, MAT2013-46308, CAIB and FEDER. M. S-B. and 
F.~D-A. thank the Theoretical Physics Group of the University of Warwick 
for the warm hospitality. 

\bibliography{references_rev}

\end{document}